\documentclass[submission,copyright,creativecommons]{eptcs}
\providecommand{\event}{ACL2 Workshop 2025} 

\usepackage{iftex,hyperref}
\usepackage[cmex10]{amsmath}
\usepackage{array}
\usepackage{url}
\usepackage{times}
\usepackage{cleveref}
\usepackage{lstautogobble}
\usepackage[dvipsnames,table]{xcolor}
\usepackage{dsfont}
\usepackage{xspace}
\usepackage[T1]{fontenc}
\usepackage[scaled]{beramono}

\newcommand{\acs}[0]{\textsc{ACL2s}\xspace}
\newcommand{\acl}[0]{\textsc{ACL2}\xspace}

\newcommand{\ie}{\emph{i.e.}}

\usepackage{makecell}

\newcommand{\bn}{Broadcastnet\xspace}
\newcommand{\fn}{Floodnet\xspace}

\makeatletter
\let\sv@thm\@thm
\def\@thm{\let\indent\relax\sv@thm}
\makeatother

\usepackage{tikz}
\usetikzlibrary{positioning, shapes.geometric, backgrounds}
\usepackage{tikzscale}

\usepackage{mathtools}

\usepackage{enumerate}
\usepackage[shortlabels]{enumitem}

\usepackage{physics}
\usepackage{tikz}
\usepackage{mathdots}
\usepackage{yhmath}
\usepackage{cancel}
\usepackage{color}
\usepackage{siunitx}
\usepackage{array}
\usepackage{multirow}
\usepackage{amssymb}
\usepackage{gensymb}
\usepackage{tabularx}
\usepackage{extarrows}
\usepackage{booktabs}
\usetikzlibrary{fadings}
\usetikzlibrary{patterns}
\usetikzlibrary{shadows.blur}
\usetikzlibrary{shapes}

\usepackage{adjustbox}
\usepackage{float}
\usepackage{pgfplots}
\pgfplotsset{compat=1.8}

\usepackage{derivative}

\newcommand*{\funcfont}{\fontfamily{lmss}\selectfont}
\newcommand*{\codefont}{\ttfamily\small}

\DeclareTextFontCommand{\funcfontify}{\funcfont}
\DeclareTextFontCommand{\codefontify}{\codefont}
\newcommand{\codify}[1]{\ensuremath{\mbox{\codefontify{#1}}}}

\newcommand{\sst}{\codify{s}\xspace}
\newcommand{\ust}{\codify{u}\xspace}

\lstdefinelanguage{acl2s}{
  keywords={defdata,definecd,definec,defun-sk,defund,defun,defthm,lambda,property,record,list,cons,match,alistof,enum,listof,alistof,range,sig,mget,mset,map,map*,b*,reduce,reduce*,<,b*,+,-,if,=,==,endp,car,cdr,caar,cadr,cdar,cddr,when,unless,!=,nil,t,<<,let,let*,cond,in,nin},
  alsoletter={-},
  otherkeywords={^,!=,==,:hyps,:name,:h,:ic,:oc,:fixed-vars,defdata-alias,defdata-subtype,create-map*,create-reduce*,check=},
  comment=[l]{;;},
  morekeywords={:b,b*},
  alsoletter={:,-,*}
}

\ifpdf
  \usepackage{underscore}         
  \usepackage[T1]{fontenc}        
\else
  \usepackage{breakurl}           
\fi

\title{A Formalization of the Correctness of the Floodsub Protocol}
\author{
Ankit Kumar\quad Panagiotis Manolios
\institute{Northeastern University\\
Boston, USA}
\email{ \{kumar.anki,p.manolios\}@northeastern.edu}
}

\newcommand{\titlerunning}{A Formalization of the Correctness of the Floodsub Protocol}
\newcommand{\authorrunning}{A. Kumar \& P. Manolios}

\hypersetup{
  bookmarksnumbered,
  pdftitle    = {\titlerunning},
  pdfauthor   = {\authorrunning},
  pdfsubject  = {EPTCS},               
}
\begin{document}
\maketitle

\begin{abstract}
  Floodsub is a simple, robust and popular peer-to-peer
  publish/subscribe (pubsub) protocol, where nodes can arbitrarily
  leave or join the network, subscribe to or unsubscribe from topics
  and forward newly received messages to all of their neighbors,
  except the sender or the originating peer. To show the correctness
  of Floodsub, we propose its specification: Broadcastsub, in which
  implementation details like network connections and neighbor
  subscriptions are elided. To show that Floodsub does really
  implement Broadcastsub, one would have to show that the two systems
  have related infinite computations. We prove this by reasoning
  locally about states and their successors using Well-Founded
  Simulation (WFS). In this paper, we focus on the mechanization of a
  proof which shows that Floodsub is a simulation refinement of
  Broadcastsub using WFS. To the best of our knowledge, ours is the
  first mechanized refinement-based verification of a real world
  pubsub protocol.
\end{abstract}

\section{Introduction}
\noindent
Peer-to-Peer (P2P) systems are decentralized distributed systems,
which constitute overlay networks built over physical networks, such
as the Internet~\cite{p2p-survey}. These systems are characterized by
self-organization, being able to handle highly dynamic network
configurations, with nodes being able to join or leave the overlay
network, which allows for scalability in the size of the networks.
Publish/Subscribe (\emph{pubsub}) systems are P2P systems that allow
(1) consumers of information (subscriber nodes) to query the system,
and (2) producers of information (publisher nodes) to publish
information to the system. Publishers are able to send messages to
multiple recipients without them having to know who the subscribers
are. This is achieved by associating subscriptions and messages with
\emph{topics}. For example, in a chat room application, each room is a
pubsub topic and clients post chat messages to rooms, which are
received by all other clients (subscribers) in the room. The Scribe
system~\cite{castro2002scribe} was a first attempt at providing
topic-based pubsub functionality over the Pastry P2P
network~\cite{pastry-p2p-location}.

A most basic implementation of pubsub systems is
\emph{Floodsub}~\cite{floodsub,libp2p-overview}. In Floodsub, nodes
are free to leave or join the network. Every node has information
about its neighboring nodes and their subscriptions. Whenever a node
joins, subscribes to or unsubscribes from a topic, it updates its
neighboring nodes. Messages are forwarded to all neighbors that
subscribe to the topic of the message, except the source (originating
node) of the message, or the node that forwarded this message. Our
implementation of Floodsub is based on its
specification~\cite{floodsub}. However, we did not do any conformance
testing or cross validation against existing implementations.  And we
do not model Ambient Peer Discovery, which is a way for nodes to learn
about their neighbors, and is described in the specification as being
external to the protocol. Given our model of Floodsub protocol, how
can we verify that it actually is an implementation of a pubsub
system? We need a specification for a pubsub system and some notion of
correctness.

We propose \emph{Broadcastsub}~\cite{libp2p-overview} as the
specification for a P2P pubsub system. Broadcastsub nodes can freely
leave and join the network. Nodes maintain a list of topics they
subscribe to. However, there is no notion of neighboring
nodes. Messages are broadcasted ``magically'' to all the subscribers
in a single transition. Notice that Broadcastsub is the simplest P2P
pubsub system, where implementation details like subscription updates,
neighboring nodes and their subscriptions are abstracted. In this
paper, we focus on the mechanization of the proof that Floodsub is a
simulation refinement~\cite{milner71} of Broadcastsub using
Well-Founded Simulation (WFS)~\cite{pete-dissertation} in \acl Sedan
(\acs)~\cite{dillinger-acl2-sedan, acl2s11}.

To show that \fn implements \bn, we will prove that \fn is a
simulation refinement of \bn. Why are we proving a simulation
refinement? Because we are comparing two P2P systems at different
levels of abstraction. In a Broadcastsub network, a message broadcast
propagates to all subscribers (of the message) instantly. However, in
a Floodsub network, a message may require several hops from one node
to another, until it reaches all of the subscribers. It is often the
case that a lower-level implementation takes several steps to match a
step of its higher level specification. Proving a WFS guarantees that
Floodsub states and related Broadcastsub states have related
computations. This notion of correctness implies that the two systems
satisfy the same $ACTL^*\setminus{X}$~\cite{bcg88} properties. WFS proofs are
structural and local, requiring proofs about states and their
successors, instead of infinite paths, thereby allowing proofs to be
amenable to formal verification. This work is a piece of a larger
puzzle that allows us to reason about more complex P2P systems using
compositional
refinement~\cite{DBLP:journals/tvlsi/ManoliosS08,DBLP:conf/date/ManoliosS04},
which we want to extend all the way down to
Gossipsub~\cite{gsreadme}. Since our models are public, protocol
engineers will be able to easily define/extend their own P2P systems
and attempt to show that their model is a refinement of one of our
existing ones. Our proof of refinement spells out exactly how the
proof breaks, if these conditions are not satisfied. Hence, our
contribution can also be used to tag P2P systems with the kinds of
network attacks they are prone to, corresponding to the refinement
conditions that were not satisfied.

We make the following contributions: (1) Formal, executable, open and
public models of Floodsub and Broadcastsub protocols expressed as
transition systems, and (2) a mechanized proof in \acs showing that
Floodsub is a simulation refinement of Broadcastsub. We discuss
related work in Section~\ref{sec:related}; and ours is the first
mechanized refinement-based verification of a real world pubsub
protocol. While refinement is a standard formal method, it has never
been previously applied to P2P pubsub protocols like FloodSub. Our
models and proofs are publicly available in our
repository~\cite{floodsubref}. Overall, our code consists of 476
theorems proved, and 10277 lines of lisp code.

\noindent
\textbf{Paper Outline.} Section~\ref{sec:model} describes \fn and \bn
models for Floodsub and Broadcastsub,
respectively. Section~\ref{sec:theorem} describes the refinement
theorem. Section~\ref{sec:organization} is a discussion about the
theorem proving process and effort that went into this
proof. Section~\ref{sec:related} discusses related
work. Section~\ref{sec:conclusion} concludes.

\section{Model Descriptions}
\label{sec:model}
We model Floodsub and its specification Broadcastsub using transition
systems consisting of states and transition relations (boolean
functions on 2 states) that depend on transition functions. We call
our models \fn and \bn respectively. In this section we will explain
each of our transition system models in a top-down fashion. The state
models are self explanatory. The interesting parts of the following
code listings are the transition relations, where we place conditions,
not only as sanity checks or for cases, but also as guards to disallow
illegal behaviour, like requiring that a peer leaving a network is
already in the network. We will discuss such conditions in detail.

\subsection{Broadcastnet}
The state of a \bn is stored in a map from peers to their
corresponding peer-states. We represent peers by natural numbers. A
\bn peer state is a record consisting of (i) \codify{pubs} : the set
of topics in which a peer publishes, (ii) \codify{subs} : the set of
topics to which a peer subscribes to, and (iii) \codify{seen} : the
set of messages a peer has already processed. We use sorted \acl lists
containing unique elements to represent sets. Hence, set equality is
reduced to list equality. Messages are records consisting of (i)
\codify{pld} : a message payload of type string (ii) \codify{tp} : the
topic in which this message was published, and (iii) \codify{or} : the
originating peer for this message.

\begin{lstlisting}
(defdata s-bn (map peer ps-bn))

(defdata-alias peer nat)

(defdata ps-bn (record
               (pubs . lot)
               (subs . lot)
               (seen . lom)))

(defdata lot (listof topic))
               
(defdata-alias topic var)

(defdata lom (listof mssg))

(defdata mssg (record
               (pld . string)
               (tp . topic)
               (or . peer)))
\end{lstlisting}

We will now define the transition relation for \bn. The relation
\codify{rel-step-bn} relates 2 \bn states \sst and \ust iff \sst
transitions to \ust. \codify{rel-step-bn} is an \codify{OR} of all the
possible ways \sst can transition to \ust. \codify{rel-skip-bn}
represents a transition where \sst chooses to skip, hence \ust is
\sst.

\begin{lstlisting}
(definec rel-step-bn (s u :s-bn) :bool
  (v (rel-skip-bn s u)
     (rel-broadcast-bn s u)
     (rel-broadcast-partial-bn s u)
     (rel-subscribe-bn s u)
     (rel-unsubscribe-bn s u)
     (rel-leave-bn s u)
     (rel-join-bn s u)))

(definecd rel-skip-bn (s u :s-bn) :bool
  (== u s))
\end{lstlisting}

\codify{rel-broadcast-bn} defines a relation between \sst and \ust,
where \ust represents the state resulting from broadcasting a message
in \sst. The broadcast is modeled as an atomic operation in which all
subscribers receive the message simultaneously. The function
\codify{(br-mssg-witness s u)} is a witness finding function that
calculates the message that was broadcast, if one exists. Since
\codify{seen} is a set of messages implemented as an ordered list with
unique elements, \codify{br-mssg-witness} utilizes this ordering of
unique messages to find the broadcasted message.

The boolean function \codify{broadcast-bn-pre} is a conjunction of the
following preconditions: (i) the broadcast message is new, \ie, it is
not already found in the \codify{seen} set of any of the peers in
\sst, (ii) the originating peer of the message exists in \sst, and
(iii) the topic of the broadcast message is one in which the
originating peer publishes messages. \codify{broadcast-bn-pre} also
appears as an input contract (\codify{:ic}) in the definition of the
\codify{broadcast} transition function. \codify{(== u (broadcast
  (br-mssg-witness s u) s))} in the definition of
\codify{rel-broadcast-bn} ensures that the message found by
\codify{br-mssg-witness} was the sole message broadcast in \sst.  We
use the \codify{insert-unique} function within \codify{broadcast-help}
to add new messages while preserving order and uniqueness of the
\codify{seen} set.

In the following code snippets, we use some \acs syntax described as
follows. \codify{\^}, \codify{v} and \codify{!} are macros for
\codify{and}, \codify{or} and \codify{not} respectively. In a
\codify{property} form, the form following the keyword \codify{:h} is
the hypothesis, while the form following the keyword \codify{:b} is
the body. \codify{(nin a x)} stands for \codify{(not (in a
  x))}. \codify{:match} is a powerful \acs pattern matching capability
which supports predicates, including recognizers automatically
generated by \codify{defdata}, disjunctive patterns and patterns
containing arbitrary code~\cite{manolios2023reasoning}. For more
expressive pattern matching, \codify{!} is used for literal match
while \codify{\&} is used as a wildcard. In a function definition form,
\codify{:ic} and \codify{:oc} abbreviate \codify{:input-contract} and
\codify{:output-contract} respectively.

\begin{lstlisting}
(definecd rel-broadcast-bn (s u :s-bn) :bool
  (^ (br-mssg-witness s u)
     (broadcast-bn-pre (br-mssg-witness s u) s)
     (== u (broadcast (br-mssg-witness s u) s))))

(definec broadcast-bn-pre (m :mssg s :s-bn) :bool
  (b* ((origin (mget :or m))
       (origin-st (mget origin s)))
    (^ (new-bn-mssgp m s)
       origin-st
       (in (mget :tp m)
           (mget :pubs origin-st)))))

(definec br-mssg-witness (s u :s-bn) :maybe-mssg
  (cond
   ((v (endp s) (endp u)) nil)
   ((== (car s) (car u)) (br-mssg-witness (cdr s) (cdr u)))
   (t (car (set-difference-equal (mget :seen (cdar u))
                                 (mget :seen (cdar s)))))))

(defdata maybe-mssg (v nil mssg))
                                   
(definecd new-bn-mssgp (m :mssg s :s-bn) :bool
  (v (endp s)
     (^ (nin m (mget :seen (cdar s)))
        (new-bn-mssgp m (cdr s)))))

(definecd broadcast (m :mssg s :s-bn) :s-bn
  :ic (broadcast-bn-pre m s)
  (broadcast-help m s))

(definecd broadcast-help (m :mssg st :s-bn) :s-bn
  (match st
    (() nil)
    (((p . pst) . rst)
     (cons `(,p . ,(if (v (in (mget :tp m) (mget :subs pst))
                          (== p (mget :or m)))
                       (mset :seen
                             (insert-unique m (mget :seen pst))
                             pst)
                     pst))
           (broadcast-help m rst)))))

(definec insert-unique (a :all x :tl) :tl
  (match x
    (() (list a))
    ((!a . &) x)
    ((e . es) (if (<< a e) (cons a x) (cons e (insert-unique a es))))))
\end{lstlisting}

We will explain \codify{rel-broadcast-partial-bn}, and its necessity
when discussing the proof of correctness later in the
paper. \codify{rel-subscribe-bn} and \codify{rel-unsubscribe-bn}
relate states \sst and \ust where \ust represents the state obtained
after a peer in \sst subscribes to or unsubscribes from a set of
topics, respectively. \codify{bn-topics-witness} calculates the peer
and the set of topics it subcribes to or unsubscribes from, if there
exists such peer. Notice that we reuse \codify{bn-topics-witness} in
the definition of \codify{rel-unsubscribe-bn}, with the arguments
reversed, so as to find the topics that are subscribed to in \sst, but
not in \ust.  The calculated set of topics are unioned with or removed
from the existing set of peer topic subscriptions of the calculated
peer, based on whether it is subscribing or unsubscribing. The
definition of \codify{unsubscribe-bn} is analogous to that of
\codify{subscribe-bn} and is hence omitted.

\begin{lstlisting}
(definecd rel-subscribe-bn (s u :s-bn) :bool
  (^ (bn-topics-witness s u)
     (mget (car (bn-topics-witness s u)) s)
     (== u (subscribe-bn (car (bn-topics-witness s u))
                         (cdr (bn-topics-witness s u))
                         s))))

(definecd rel-unsubscribe-bn (s u :s-bn) :bool
  (^ (bn-topics-witness u s)
     (mget (car (bn-topics-witness u s)) s)
     (== u (unsubscribe-bn (car (bn-topics-witness u s))
                           (cdr (bn-topics-witness u s))
                           s))))

(definec bn-topics-witness (s u :s-bn) :maybe-ptops
  (cond
   ((v (endp s) (endp u)) nil)
   ((== (car s) (car u)) (bn-topics-witness (cdr s) (cdr u)))
   ((^ (== (caar s) (caar u))
       (set-difference-equal (mget :subs (cdar u))
                             (mget :subs (cdar s))))
    (cons (caar s)
          (set-difference-equal (mget :subs (cdar u))
                                (mget :subs (cdar s)))))
   (t nil)))

(defdata maybe-ptops (v nil (cons peer lot)))

(definecd subscribe-bn (p :peer topics :lot s :s-bn) :s-bn
  :ic (mget p s)
  (let ((pst (mget p s)))
    (mset p (mset :subs (union-equal (mget :subs pst) topics) pst) s)))
\end{lstlisting}

\codify{rel-join-bn} and \codify{rel-leave-bn} relate states \sst and
\ust where \ust is obtained after a peer joins \sst or leaves \sst,
respectively. \codify{bn-join-witness} calculates the peer and its
peer-state, if there exists a peer that joins \sst. Its definition
depends on the keys of our \bn state being in order, which is
guaranteed by \acs \codify{map}s. A peer joins a \bn state when a new
default \bn peer state is set for the corresponding peer. A peer
leaves a \bn state when the entry corresponding to the leaving peer is
removed from the state.

\begin{lstlisting}
(definecd rel-join-bn (s u :s-bn) :bool
  (^ (bn-join-witness s u)
     (b* ((p (car (bn-join-witness s u)))
          (pst (cdr (bn-join-witness s u))))
       (^ (! (mget p s))
          (== u (join-bn p (mget :pubs pst) (mget :subs pst) s))))))

(definecd rel-leave-bn (s u :s-bn) :bool
  (^ (bn-join-witness u s)
     (mget (car (bn-join-witness u s)) s)
     (== u (leave-bn (car (bn-join-witness u s)) s))))

(definec bn-join-witness (s u :s-bn) :maybe-ppsbn
  (match (list s u)
    ((() ((q . qst) . &)) `(,q . ,qst))
    ((((p . pst) . rs1) ((q . qst) . rs2))
     (cond
      ((== `(,p . ,pst) `(,q . ,qst)) (bn-join-witness rs1 rs2))
      ((!= q p) `(,q . ,qst))  ;; Joining peer found
      (t nil)))
    (& nil)))

(defdata maybe-ppsbn (v nil (cons peer ps-bn)))

(definecd join-bn (p :peer pubs subs :lot s :s-bn) :s-bn
  :ic (! (mget p s)) ;; Join only if peer does not already exist in state
  (mset p (ps-bn pubs subs '()) s))

(definecd leave-bn (p :peer s :s-bn) :s-bn
  :ic (mget p s) ;; Leave only if peer already exists in state
  (match s
    (((!p . &) . rst) rst)
    ((r . rst) (cons r (leave-bn p rst)))))
\end{lstlisting}

\subsection{Floodnet}
A \fn peer-state is a record consisting of sets \codify{pubs},
\codify{subs} and \codify{seen} which we described previously in
context of \bn peer-states. It also consists of \codify{pending},
which is a set of messages that have not yet been processed, and
\codify{nsubs}, a map from topics to list of peers. \codify{nsubs}
stores topic subscriptions for neighboring peers.
\begin{lstlisting}
(defdata s-fn (map peer ps-fn))

(defdata ps-fn
  (record (pubs . lot)
          (subs . lot)
          (nsubs . topic-lop-map)
          (pending . lom)
          (seen . lom)))
\end{lstlisting}

When a message has not been forwarded to neighboring subscribers
(processed) it remains in the \codify{pending} set. Once it is
processed, it is added to the \codify{seen} set. In our \fn model,
\codify{pending} and \codify{seen} are sets of messages, instead of
queues. This simplifies the model and allows us to not worry about the
order in which messages are received. Related states have equal sets
of \codify{seen} messages.

We define the transition relation \codify{rel-step-fn} which relates
two \fn states \sst and \ust iff \sst transitions to \ust. It encodes
all the possible ways \sst can transition to \ust. \codify{rel-skip-fn}
represents a transition where \sst chooses to skip, hence \ust is
\sst.

\begin{lstlisting}[label=lst:rel-step-fn]
(definec rel-step-fn (s u :s-fn) :bool
  (v (rel-skip-fn s u)
     (rel-produce-fn s u)
     (rel-forward-fn s u)
     (rel-subscribe-fn s u)
     (rel-unsubscribe-fn s u)
     (rel-leave-fn s u)
     (rel-join-fn s u)))

(definecd rel-skip-fn (s u :s-fn) :bool
  (== u s))
\end{lstlisting}

\codify{rel-produce-fn} relates \sst and \ust where \ust represents the state
obtained after a new message has been produced in \sst. The newly produced
message is one of the pending messages in \ust. The boolean function
\codify{produce-fn-pre} is a conjunction of the following
preconditions: (i) the produced message is new \ie, it is not already
found in the \codify{seen} or \codify{pending} sets of any of the
peers in \sst, (ii) the originating peer of the message exists in
\sst, and (iii) the topic of the produced message is one in which the
originating peer publishes messages. The new message is added to the
set of pending messages of the originating peer.

\begin{lstlisting}
(definecd rel-produce-fn (s u :s-fn) :bool
  (rel-produce-help-fn s u (fn-pending-mssgs u)))

(definec fn-pending-mssgs (s :s-fn) :lom
  (match s
    (() '())
    (((& . pst) . rst) (union-set (mget :pending pst) (fn-pending-mssgs rst)))))  

(definec rel-produce-help-fn (s u :s-fn ms :lom) :bool
  (match ms
    (() nil)
    ((m . rst) (v (^ (produce-fn-pre m s)
                     (== u (produce-fn m s)))
                  (rel-produce-help-fn s u rst)))))

(definec produce-fn-pre (m :mssg s :s-fn) :bool
  (b* ((origin (mget :or m))
       (origin-st (mget origin s)))
    (^ (new-fn-mssgp m s)
       origin-st
       (in (mget :tp m) (mget :pubs origin-st)))))

(definecd new-fn-mssgp (m :mssg s :s-fn) :bool
  (v (endp s)
     (^ (nin m (mget :seen (cdar s)))
        (nin m (mget :pending (cdar s)))
        (new-fn-mssgp m (cdr s)))))

(definecd produce-fn (m :mssg s :s-fn) :s-fn
  :ic (produce-fn-pre m s)
  (mset (mget :or m)
        (add-pending-psfn m (mget (mget :or m) s)) s))

(definecd add-pending-psfn (m :mssg pst :ps-fn) :ps-fn
  (if (v (in m (mget :pending pst))
         (in m (mget :seen pst)))
      pst
    (mset :pending (cons m (mget :pending pst)) pst)))
\end{lstlisting}

\codify{rel-forward-fn} relates states \sst and \ust where \ust
represents the state obtained after a peer in \sst forwards a pending
message. Notice that any of the pending messages in \sst are eligible
to be forwarded. Notice also that there can be several peers with a
given message pending, and the \fn can take several possible
transitions to states related to the current state by
\codify{rel-forward-fn}. To model this in a constructive and
deterministic way, we introduce \codify{find-forwarder} as a skolem
function which returns the first peer in the state where a given
message is pending. It produces a concrete peer $p$ in the call to
\codify{forward-fn}. Its output contract (\codify{:oc}) specifies that
the message forwarding peer it returns (i) is a peer in \sst, (ii)
possesses the given message in its \codify{pending} set, and (iii)
that message is not new in \sst.

The \codify{forward-fn} transition function simultaneously updated the
state of the peer that forwards the message, using
\codify{update-forwarder-fn} and updates the \codify{pending} sets of
the neighboring subscribers by inserting the forwarded message using
\codify{forward-help-fn}. Note that messages are forwarded to all the
peers subscribing to the topic of the message in the \codify{:nsubs}
map. If the forwarding peer records its own subscriptions in
\codify{:nsubs}, it can lead to \codify{rel-forward-fn} being
infinitely enabled. We will ensure that a peer does not include itself
in this map, by considering \emph{good} \fn states later in the paper.

\begin{lstlisting}[label=lst:forward-fn]
(definecd rel-forward-fn (s u :s-fn) :bool
    (rel-forward-help-fn s u (fn-pending-mssgs s)))

(definec rel-forward-help-fn (s u :s-fn ms :lom) :bool
  (match ms
    (() nil)
    ((m . rst)
     (v (^ (in m (fn-pending-mssgs s))
           (== u (forward-fn (find-forwarder s m) m s)))
        (rel-forward-help-fn s u rst)))))

(definec find-forwarder (s :s-fn m :mssg) :peer
    :ic (in m (fn-pending-mssgs s))
    :oc (^ (mget (find-forwarder s m) s)
           (in m (mget :pending (mget (find-forwarder s m) s)))
           (! (new-fn-mssgp m s)))
    (match s
      (((p . &)) p)
      (((p . pst) . rst)
       (if (in m (mget :pending pst)) p (find-forwarder rst m)))))
          
(definecd forward-fn (p :peer m :mssg s :s-fn) :s-fn
  :ic (^ (mget p s)
         (in m (mget :pending (mget p s))))
  (b* ((tp (mssg-tp m))
       (pst (mget p s))
       (nsubs (mget :nsubs pst))
       (fwdnbrs (mget tp nsubs)))
    (forward-help-fn (update-forwarder-fn p m s) fwdnbrs m)))

(definec update-forwarder-fn (p :peer m :mssg s :s-fn) :s-fn
  (match s
    (() '())
    (((!p . pst) . rst) (cons `(,p . ,(forwarder-new-pst pst m)) rst))
    ((r . rst) (cons r (update-forwarder-fn p m rst)))))

(definecd forwarder-new-pst (pst :ps-fn m :mssg) :ps-fn
  (mset :seen
        (insert-unique m (mget :seen pst))
        (mset :pending
              (remove-equal m (mget :pending pst))
              pst)))

(definecd forward-help-fn (s :s-fn nbrs :lop m :mssg) :s-fn
  (match s
    (() '())
    (((q . qst) . rst)
     (cons (if (in q nbrs)
               `(,q . ,(add-pending-psfn m qst))
             `(,q . ,qst))
           (forward-help-fn rst nbrs m)))))
\end{lstlisting}

\codify{rel-subscribe-fn} and \codify{rel-unsubscribe-fn} relate
states \sst and \ust where \ust represents the state obtained after a
peer in \sst subscribes to or unsubscribes from a set of topics,
respectively. They are very similar to their \bn counterparts and
hence we omit their definitions.

\codify{rel-join-fn} and \codify{rel-leave-fn} relate states \sst and
\ust where \ust represents the state obtained after a peer joins \sst
or leaves \sst, respectively. \codify{fn-join-witness} calculates the
peer and its peer-state, if there exists a peer that joins \sst, and
is analogous to \codify{bn-join-witness}. \codify{rel-join-fn}
requires that the joining peer (i) is not already in \sst, and (ii)
does not exist in its own \codify{:nsubs} map. The second condition is
necessary to prevent peers from endlessly forwarding messages to
themselves. \codify{join-fn} depends on \codify{new-joinee-st-fn}
which returns the state for a newly joined peer, and on
\codify{set-subs-sfn} which updates the \codify{:nsubs} map for each
of the neighboring peers of the joining node. For the sake of brevity,
we omit the definitions of these helper functions. The
\codify{rel-leave-fn} relation, similar to \codify{rel-leave-bn}
requires that there is a leaving peer, as calculated by
\codify{(fn-join-witness u s)} and that it already exist in the
state. Notice that there is another requirement, that a leaving peer
has no pending messages. This condition allows for graceful exit of
leaving peers, guaranteeing that no pending messages are lost along
with them.

\begin{lstlisting}[label=lst:join-leave-fn]
(definecd rel-join-fn (s u :s-fn) :bool
  (^ (fn-join-witness s u)
     (b* ((p (car (fn-join-witness s u)))
          (pst (cdr (fn-join-witness s u)))
          (nbrs (topic-lop-map->lop (mget :nsubs pst))))
       (^ (! (mget p s))
          (nin p nbrs)
          (== u (join-fn p (mget :pubs pst) (mget :subs pst) nbrs s))))))

(definecd join-fn (p :peer pubs subs :lot nbrs :lop s :s-fn) :s-fn
  :ic (^ (! (mget p s))
	       (nin p nbrs))
  (set-subs-sfn nbrs
                subs
                p
                (mset p (new-joinee-st-fn pubs subs nbrs s) s)))

(definecd rel-leave-fn (s u :s-fn) :bool
  (^ (fn-join-witness u s)
     (mget (car (fn-join-witness u s)) s)
     (endp (mget :pending (mget (car (fn-join-witness u s)) s)))
     (== u (leave-fn (car (fn-join-witness u s)) s))))

(definecd leave-fn (p :peer s :s-fn) :s-fn
  :ic (mget p s)
  (match s
    (() '())
    (((!p . &) . rst) rst)
    ((r . rst) (cons r (leave-fn p rst)))))
\end{lstlisting}

\section{Correctness and the Refinement Theorem}
\label{sec:theorem}
We consider simulation refinement~\cite{pete-dissertation} as the
notion of correctness for \fn and show that \fn is a simulation
refinement of \bn. The key idea of a simulation refinement is to show
that every behavior of the concrete system (\fn) is allowed by the
abstract system (\bn). If we prove a WFS refinement then we know that
for any infinite computation tree starting from some \fn state, we can
find a related computation tree in \bn after applying the refinement
map.  Another consequence is that we preserve any branching time
properties, excluding next time, for example, all properties in
$ACTL^*\setminus X$~\cite{bcg88}.

The refinement map needs to map \fn states to ``related'' \bn states.
Why do we require a refinement map? Because states in different levels
of abstractions may represent data differently, or some implementation
details from the lower abstraction may simply be missing in the higher
level specification. For example, \codify{:nsubs} and
\codify{:pending} appear only in \fn peer states, not in \bn peer
states. Using a refinement map is like putting on glasses that let us
``see” lower-level concrete states as their corresponding abstract
specification states. The refinement map that we use is \codify{f2b}
shown below. It maps \fn states to \bn states where pending messages
have not yet been broadcasted. This is called as the commitment
approach to refinement, since we are mapping to states consisting of
only those messages that have been fully propagated in \fn and are
thus considered committed.

\begin{lstlisting}[label=lst:f2b]
(definec f2b (s :s-fn) :s-bn
  (f2b-help s (fn-pending-mssgs s)))

(definec f2b-help (s :s-fn ms :lom) :s-bn
  (if (endp s)
      '()
    (cons `(,(caar s) . ,(f2b-st (cdar s) ms))
          (f2b-help (cdr s) ms))))

(definecd f2b-st (ps :ps-fn ms :lom) :ps-bn
  (ps-bn (mget :pubs ps)
         (mget :subs ps)
         (set-difference-equal (mget :seen ps) ms)))
\end{lstlisting}

\begin{figure}[H]
  \centering
  \begin{tikzpicture}
    \input{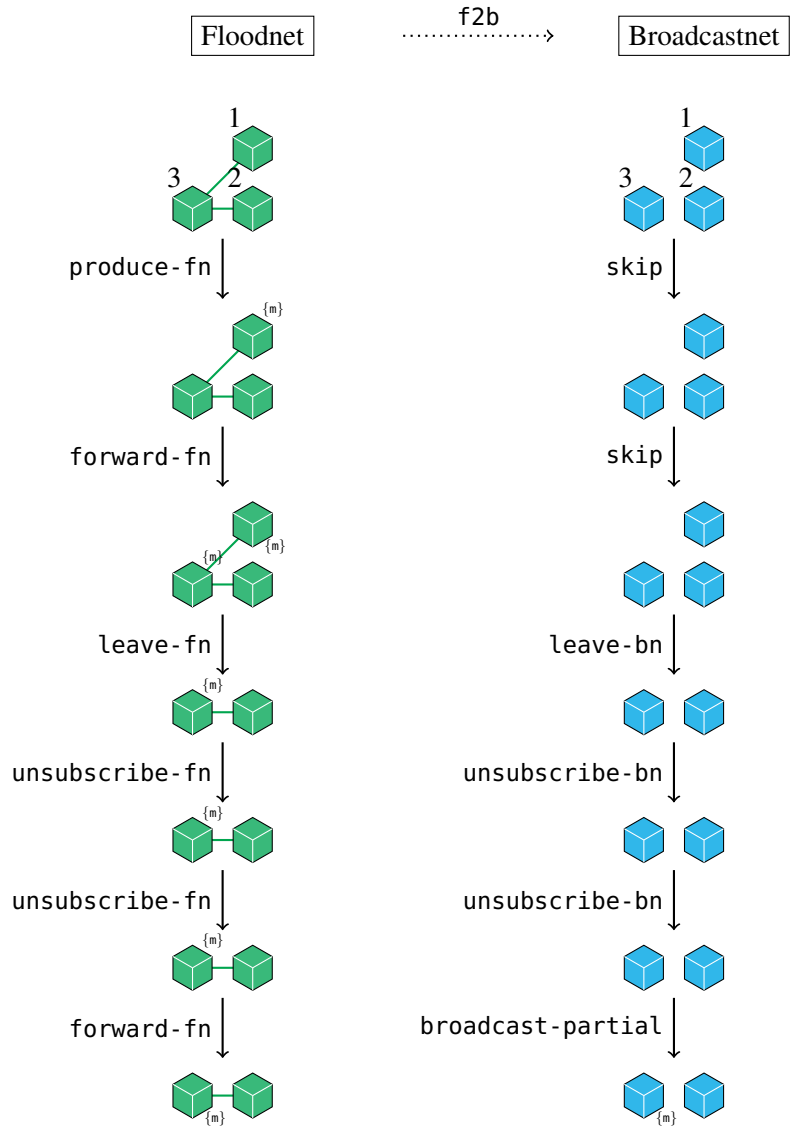};
    \node[draw] at (-3,2) {\fn};
    \node[draw] at (3,2) {\bn};
    \draw[dotted,thick,->] (-1,2) -- (1,2) node[midway, above]{\codify{f2b}};
    
    \pic at (3,0.5) {hex1={l=1}};
    \pic at (3,-0.3) {hex1={l=2}};
    \pic at (2.2,-0.3) {hex1={l=3}};
    \draw[->,thick] (2.6,-0.7) -- (2.6,-1.5) node[midway, left] {\codify{skip}};
    \pic at (3,-2) {hex1};
    \pic at (3,-2.8) {hex1};
    \pic at (2.2,-2.8) {hex1};
    \draw[->,thick] (2.6,-3.2) -- (2.6,-4) node[midway, left] {\codify{skip}};
    \pic at (3,-4.5) {hex1};
    \pic at (3,-5.3) {hex1};
    \pic at (2.2,-5.3) {hex1};
    \draw[->,thick] (2.6,-5.7) -- (2.6,-6.5) node[midway, left] {\codify{leave-bn}};
    \pic at (3,-7) {hex1};
    \pic at (2.2,-7) {hex1};
    \draw[->,thick] (2.6,-7.4) -- (2.6,-8.2) node[midway, left] {\codify{unsubscribe-bn}};
    \pic at (3,-8.7) {hex1};
    \pic at (2.2,-8.7) {hex1};
    \draw[->,thick] (2.6,-9.1) -- (2.6,-9.9) node[midway, left] {\codify{unsubscribe-bn}};
    \pic at (3,-10.4) {hex1};
    \pic at (2.2,-10.4) {hex1};
    \draw[->,thick] (2.6,-10.8) -- (2.6,-11.6) node[midway, left] {\codify{broadcast-partial}};
    \pic at (3,-12.1) {hex1};
    \pic at (2.2,-12.1) {hex1={br=$\{\codify{m}\}$}};

    \draw[-,thick,Green]  (-3,0.5) -- (-3.8,-0.3);
    \draw[-,thick,Green]  (-3,-0.3) -- (-3.8,-0.3);
    \pic at (-3,0.5) {hex2={l=1}};
    \pic at (-3,-0.3) {hex2={l=2}};
    \pic at (-3.8,-0.3) {hex2={l=3}};
    \draw[->,thick] (-3.4,-0.7) -- (-3.4,-1.5) node[midway, left] {\codify{produce-fn}};
    \draw[-,thick,Green]  (-3,-2) -- (-3.8,-2.8);
    \draw[-,thick,Green]  (-3,-2.8) -- (-3.8,-2.8);
    \pic at (-3,-2) {hex2={tr=$\{\codify{m}\}$}};
    \pic at (-3,-2.8) {hex2};
    \pic at (-3.8,-2.8) {hex2};
    \draw[->,thick] (-3.4,-3.2) -- (-3.4,-4) node[midway, left] {\codify{forward-fn}};
    \draw[-,thick,Green]  (-3,-4.5) -- (-3.8,-5.3);
    \draw[-,thick,Green]  (-3,-5.3) -- (-3.8,-5.3);
    \pic at (-3,-4.5) {hex2={br=$\{\codify{m}\}$}};
    \pic at (-3,-5.3) {hex2};
    \pic at (-3.8,-5.3) {hex2={tr=$\{\codify{m}\}$}};
    \draw[->,thick] (-3.4,-5.7) -- (-3.4,-6.5) node[midway, left] {\codify{leave-fn}};
    \draw[-,thick,Green]  (-3,-7) -- (-3.8,-7);
    \pic at (-3,-7) {hex2};
    \pic at (-3.8,-7) {hex2={tr=$\{\codify{m}\}$}};
    \draw[->,thick] (-3.4,-7.4) -- (-3.4,-8.2) node[midway, left] {\codify{unsubscribe-fn}};
    \draw[-,thick,Green]  (-3,-8.7) -- (-3.8,-8.7);
    \pic at (-3,-8.7) {hex2};
    \pic at (-3.8,-8.7) {hex2={tr=$\{\codify{m}\}$}};
    \draw[->,thick] (-3.4,-9.1) -- (-3.4,-9.9) node[midway, left] {\codify{unsubscribe-fn}};
    \draw[-,thick,Green]  (-3,-10.4) -- (-3.8,-10.4);
    \pic at (-3,-10.4) {hex2};
    \pic at (-3.8,-10.4) {hex2={tr=$\{\codify{m}\}$}};
    \draw[->,thick] (-3.4,-10.8) -- (-3.4,-11.6) node[midway, left] {\codify{forward-fn}};
    \draw[-,thick,Green]  (-3,-12.1) -- (-3.8,-12.1);
    \pic at (-3,-12.1) {hex2};
    \pic at (-3.8,-12.1) {hex2={br=$\{\codify{m}\}$}};
  \end{tikzpicture}
  \caption{On the left is an example \fn trace. \bn states on the
    right are refinement maps of the \fn states on the left, and every
    step taken by the \bn states matches each step taken by the \fn
    states.}
  \label{fig:traces}
\end{figure}

To gain a better understanding of our refinement map, we examine
example traces of \fn and \bn in Figure~\ref{fig:traces}. On the left
side, we have a trace of a \fn, consisting of 3 green colored nodes,
numbered 1, 2 and 3. The node numbered 3 is connected to nodes 1 and
2. We show pending messages on the top left of a node, and seen
messages on the bottom right. So, in the second \fn state shown, node
1 has a pending message \codify{m}, after a \codify{produce-fn}
transition. On the right side, we have \bn states such that for each
\fn state on the left, we have its refinement map on the right and for
each transition on the left, we show a corresponding matching
transition on the right.

The transitions on the \fn side are as follows : (i) Node 1 produces
message \codify{m}; (ii) Node 1 forwards its pending message
\codify{m} to its connected neighboring peer 3; (iii) Node 1 leaves
the network (iv) Node 2 unsubscribes from \codify{(mssg-tp m)}, which
is the message topic (iv) Node 3 unsubscribes from \codify{(mssg-tp
  m)}, and finally (v) Node 3 forwards \codify{m} to node 2. Notice
that \codify{f2b} is a clear refinement map where events like joining
and leaving are not masked. Hence, in the corresponding \bn states,
leave and unsubscribe transitions are matched with leave and
unsubscribe transitions. However, when the message \codify{m} can no
longer be forwarded, and is no longer pending, it needs to be matched
by a \codify{broadcast}. But notice that on the \bn side (a) the
originating peer (Node 1) is no longer present, which is required for
a \codify{broadcast}, due to \codify{broadcast-bn-pre}, and (b) there
are no subscribers of \codify{m} left in the network! This issue
arises from the fact that broadcasting a message in \fn is a highly
fragmented operation, taking place over several message hops during
which peers are free to leave, join, subscribe or unsubscribe. With so
many moving parts in the network, it becomes impossible to specify
which nodes will receive the broadcasted message in the \bn under the
refinement map at the time the message is produced. To solve this
problem, we generalize the \bn specification by adding another
transition relation: \codify {rel-broadcast-partial-bn} which allows
us to relate 2 states \sst and \ust where \ust represents the state
obtained after broadcasting a message in \sst, but only
partially. This relation is defined using the
\codify{broadcast-partial} transition function, which given a message
and a list of peers, sends the message to those peers.

When we consider a static configuration where nodes do not change
their subscriptions, no existing nodes are leaving, no new nodes are
joining, and the network is connected in each topic, the recipients of
the message \codify{m} in \codify{broadcast-partial} can be shown to
be exactly those in \codify{broadcast} \ie, the subscribers of
\codify{(mssg-tp m)}. This aligns with Dijkstra's notion of
self-stabilizing distributed systems~\cite{dijkstra74}, where the
system is guaranteed to reach a legitimate configuration regardless of
the initial state. In our case, once the system stabilizes (i.e., peer
churn and subscription/unsubscription ceases), the
\codify{broadcast-partial} step effectively becomes indistinguishable
from a \codify{broadcast} step, reinforcing the view that
\codify{broadcast-partial} is a generalization that accommodates
transient perturbations while preserving desirable behavior in steady
state.

\begin{lstlisting}[label=lst:rel-bcast-partial]
(definecd rel-broadcast-partial-bn (s u :s-bn) :bool
  (^ (br-mssg-witness s u)
     (new-bn-mssgp (br-mssg-witness s u) s)
     (== u (broadcast-partial (br-mssg-witness s u)
                              (brd-receivers-bn (br-mssg-witness s u) u)
                              s))))

(definecd broadcast-partial (m :mssg ps :lop s :s-bn) :s-bn
  :ic (new-bn-mssgp m s)
  (broadcast-partial-help m ps s))

(definecd broadcast-partial-help (m :mssg ps :lop st :s-bn) :s-bn
  (match st
    (() nil)
    (((p . pst) . rst)
     (cons `(,p . ,(if (== p (car ps))
                       (mset :seen (insert-unique m (mget :seen pst)) pst)
                     pst))
           (broadcast-partial-help m (if (== p (car ps)) (cdr ps) ps) rst)))))
;; broadcast message receivers in a Broadcastnetwork
(definec brd-receivers-bn (m :mssg s :s-bn) :lop
  (match s
    (() ())
    (((p . pst) . rst) (if (in m (mget :seen pst))
                           (cons p (brd-receivers-bn m rst))
                         (brd-receivers-bn m rst)))))
\end{lstlisting}

We define \codify{rel-B}, which holds for related
states. \codify{rel-B} is defined over a set of states combining both
\fn and \bn states, which we define as \codify{borf}.  Notice that
\codify{rel-B} depends on \fn states satisfying the
\codify{good-s-fnp} predicate. This predicate ensures that each \fn
state in the trace satisfies certain invariants. Given space
constraints, we only list the invariant properties that hold true for
\codify{good-s-fnp} states at the end of the following listing.

\begin{lstlisting}
(defdata borf (v s-bn s-fn))

(definec rel-B (x y :borf) :bool
  (v (rel-wf x y)
     (== x y)))

(definec rel-wf (x y :borf) :bool
  (^ (s-fnp x)
     (s-bnp y)
     (good-s-fnp x)
     (== y (f2b x))))

(definec rel-> (s u :borf) :bool
  (v (^ (s-fnp s) (s-fnp u) (good-rel-step-fn s u))
     (^ (s-bnp s) (s-bnp u) (rel-step-bn s u))))

(definec good-rel-step-fn (s u :s-fn) :bool
  (^ (good-s-fnp s)
     (good-s-fnp u)
     (rel-step-fn s u)))

;; A good-s-fnp state satisfies 2 predicates
(definec good-s-fnp (s :s-fn) :bool
  (^ (p!in-nsubs-s-fn s) (ordered-seenp s)))

;; Invariant 1: A peer p does not track its own subscriptions in the
;; :nsubs map. So, it can not forward a message to itself.
(propertyd prop=p!in-nsubs-s-fn (p :peer tp :topic s :s-fn)
  :h (^ (mget p s) (p!in-nsubs-s-fn s))
  :b (nin p (mget tp (mget :nsubs (mget p s)))))

;; Invariant 2: :seen components of Floodnet peers are ordered.
(property prop=ordered-seenp-cdar (s :s-fn)
  :h (^ s (ordered-seenp s))
  :b (orderedp (mget :seen (cdar s))))

(definec orderedp (x :tl) :bool
  (match x
    (() t)
    ((&) t)
    ((a . (b . &)) (^ (<< a b) (orderedp (cdr x))))))
\end{lstlisting}

Proving the WFS refinement requires proving the following three
theorems: (i) WFS1 states that concrete \fn states are related to
their corresponding \bn states under the refinement map (by relation
\codify{rel-B}), (ii) WFS2 states that the labelling function labels
related states equally, and (iii) WFS3 states that given related
states \codify{s} and \codify{w}, and given \codify{s} steps to
\codify{u} under the transition relation, there exists a state, say
\codify{v}, such that \codify{u} is matched by a step from \codify{w}
going to \codify{v} such that \codify{w} is related to \codify{v}.

\begin{lstlisting}
;; WFS1
(property b-maps-f2b (s :s-fn)
  :h (good-s-fnp s)
  :b (rel-B s (f2b s)))

;; WFS2. L is the labelling functions of our combined transition system    
(definec L (s :borf) :borf
  (match s
    (:s-bn s)
    (:s-fn (f2b s))))

(property wfs2 (s w :borf)
  :h (rel-B s w)
  :b (== (L s) (L w)))

 ;; WFS3     
(defun-sk exists-v-wfs (s u w)
  (exists (v)
    (^ (rel-> w v)
       (rel-B u v))))

(property wfs3 (s w u :borf)
  :h (^ (rel-B s w)
        (rel-> s u))
  :b (exists-v-wfs s u w))

;; Witness generating function for v
(definec exists-v (s u w :borf) :borf
  :ic (^ (rel-B s w)
         (rel-> s u))
  (if (null s)
      (if (null u)
          nil
        (exists-nil-v u))
    (exists-cons-v s u w)))

;; when w is nil
(definec exists-nil-v (u :borf) :borf
  :ic (^ u (rel-> nil u))
  (match u
    (:s-fn (exists-v1 nil u))
    (:s-bn u)))

;; when w is not nil
(definec exists-cons-v (s u w :borf) :borf
  :ic (^ s (rel-B s w) (rel-> s u))
  (cond
   ((^ (s-bnp s) (s-bnp w)) u)
   ((^ (s-fnp s) (s-bnp w)) (exists-v1 s u))
   ((^ (s-fnp s) (s-fnp w)) u)))

;; when s and u are Floodnet states
(definec exists-v1 (s u :s-fn) :s-bn
  :ic (good-s-fnp s)
  (cond
   ((rel-skip-fn s u) (f2b s))
   ((^ (rel-forward-fn s u)
       (!= (f2b s) (f2b u)))
    (broadcast-partial (br-mssg-witness (f2b s) (f2b u))
                       (brd-receivers-bn (br-mssg-witness (f2b s) (f2b u))
                                         (f2b u))
                       (f2b s)))
   (t (f2b u))))
\end{lstlisting}

\section{Proof Organization}
\label{sec:organization}
Given that we define our models using transition functions from states
to states, whereas the refinement theorem is expressed in terms of
transition relations, proving the monolithic refinement theorem can be
a daunting task. In this section, we describe how we approached the
mechanization of the refinement proof.  The entire codebase can be
logically partitioned into four stages:
\begin{itemize}
\item \textbf{State models and transition functions}: State models are
  described using \codify{defdata}, and transition functions on the
  state models are described using \codify{definec} and appear in
  files \codify{bn-trx.lisp} and \codify{fn-trx.lisp}. Apart from the
  input state, functions may accept additional arguments. For example
  \codify{forward-fn}~\ref{lst:forward-fn} accepts a peer \codify{p}
  along with a message \codify{m} that \codify{p} forwards. These
  functions usually have input contracts to ensure that the extra
  arguments satisfy certain properties, for example, \codify{p} should
  be a peer in the state \codify{s}, and it should have \codify{m} in
  its \codify{pending} set of messages.
\item \textbf{Properties of functions under the refinement map}: Given that we
  have defined functions that accept states and output states, we then
  prove theorems relating \fn states to their corresponding \bn states
  under the refinement map in file \codify{f2b-commit.lisp}. For
  example, here is one such theorem:
    \begin{lstlisting}
      (property prop=forward-fn (p :peer m :mssg s :s-fn)
       :h (^ (mget p s)
             (in m (mget :pending (mget p s)))
             (== (fn-pending-mssgs (forward-fn p m s))
                 (fn-pending-mssgs s)))
       :b (== (f2b (forward-fn p m s))
              (f2b s)))
     \end{lstlisting}
     Notice that these theorems still depends on variables
     that have not yet been skolemized, so as to ease the theorem
     proving process.
   \item \textbf{Transition relations}: We define the transitions
     relations for both \bn and \fn in
     \codify{trx-rels.lisp}. Transition relations are boolean
     functions over two state variables, and hence, we also define
     witness functions for non-state variables appearing in the
     transition functions. In our running example, we instantiate
     \codify{p} with \codify{(find-forwarder s m)} and \codify{m}
     could be any one of the pending messages in the state.
   \item \textbf{Combined states, transitions relations and
       correctness theorems}: Finally we prove the WFS theorems and
     their helper properties in \codify{f2b-sim-ref.lisp}.
\end{itemize}

During development, some of the previous iterations of our model
deviated from the metatheory. For example, in one iteration of the
final theorem, the restrictions on the transition relations, which
serve as guards against illegal behaviors, emerged as part of the
hypotheses of WFS3. It forced us to understand the nature of good
states, and to derive the required hypotheses from the invariants of
the good states. In an another iteration of our models, the transition
relations corresponding to each of the transitions a model can make,
was augmented with natural numbers, such that transitions on the \fn
states were matched by transitions on the \bn states bearing the same
number. Eventually, this arrangement seemed unnecessary because even
without the natural numbers, the theorem prover was able to pick the
required transition based on theorems proved on them, and because of
their definitions being disabled. Hence we would recommend to stick to
the metatheory when implementing proofs of refinement, and always
write the top level theorems, before embarking on proving lower-level
theorems.

\section{Related Work}
\label{sec:related}
Proof mechanization in context of P2P systems has been explored
previously. Azmy et. al.~\cite{pastryCorrect} formally verified a
safety property of Pastry, a P2P Distributed Hash Table (DHT), in
TLA+~\cite{lamport2002specifying}. The safety property is that of
correct delivery, which states that at any point in time, there is at
most one node that answers a lookup request for a key, and this node
must be the closest live node to that key. Their proof assumes that
nodes never fail, which is a likely event in any P2P
system. Zave~\cite{chordZave} utilized the Alloy tool~\cite{alloy} to
produce counter-examples to show that no published version of Chord is
correct w.r.t. the liveness property of the Chord ring-maintenance
protocol: that the protocol can eventually repair all disruptions in
the ring structure, given ample time and no further disruptions while
it is working. Kumar et. al. modeled the Gossipsub~\cite{gsreadme} P2P
protocol in \acs, formulated safety properties for its scoring
function and showed using counter-examples that for some applications
like Ethereum which configure Gossipsub in a particular way, it is
possible for Sybil nodes to violate those properties, thereby creating
large scale partition or eclipse attacks on the
network~\cite{KumarGossip,KumarGWS}. To the best of our knowledge, none of the
previous work has attempted to prove the correctness of a P2P system
by showing it as a refinement of a higher level specification.

There exist several provers to formally check properties of
distributed systems, such as Dafny~\cite{hawblitzel2015ironfleet},
TLA+, Ivy~\cite{padon-ivy} and
DistAlgo~\cite{derecho}. They operate by reducing a given
specification to a decidable logic formula expressed entirely in First
Order Logic. The basic tactic involves forming a conjunction of
protocol invariants, invert it, and then using an SMT solver to
(possibly) search for a counterexample. The issue in such systems is a
lack of expressivity, which does not allow capturing properties over
infinite traces. Another issue is that our models can be arbitrary,
with nodes leaving and joining and with arbitrary pending messages in
transit across a network. And we are reasoning about all possible
behaviors of the protocol, which could not be done if we were to be
limited to a decidable fragment of logic.

We wrote our models and proved our theorems in \acs. The \acl Sedan
(\acs)~\cite{dillinger-acl2-sedan, acl2s11} is an extension of the
\acl theorem prover\cite{acl2-car, acl2-acs, acl2-web}. On top of the
capabilities of \acl, \acs provides the following: (1) A powerful
type system via the \texttt{defdata} data definition
framework~\cite{defdata} and the \texttt{definec} and
\texttt{property} forms, which support typed definitions and
properties.  (2) Counterexample generation capability via the
\texttt{cgen} framework, which is based on the synergistic integration
of theorem proving, type reasoning and testing~\cite{cgen,
  harsh-fmcad, harsh-dissertation}.  (3) A powerful termination
analysis based on calling-context graphs~\cite{ccg} and
ordinals~\cite{ManoliosVroon03, ManoliosVroon04, MV05}.  (4) An
(optional) Eclipse IDE plugin~\cite{acl2s11}.  (5) The \acs systems
programming framework (ASPF)~\cite{acl2s-systems-programming} which
enables the development of tools in Common Lisp that use ACL2, \acs
and Z3 as a
service~\cite{enumerative-data-types,invariant-discovery-game,
  ankit-mpmt, acl2-workshop-checker-paper}.

\section{Conclusions and Future Work}
\label{sec:conclusion}
In this paper, we described our \acs models for Broadcastsub and
Floodsub, \bn and \fn respectively, and proposed \bn as a
specification of \fn. For both the models, we explained our transition
systems (including state and transition relations) and design
decisions. We described our refinement map \codify{f2b}, the combined
transition system and the equivalence relation \codify{rel-B} relating
related states. Finally we explained the refinement theorem.

In the future we would like to show that in a static configuration
where a \fn is connected in each of the topics, a \codify{forward-fn}
transition can be matched by either a \codify{broadcast-partial} or a
\codify{broadcast} transition. We would also like to refine \fn
progressively until we approach a specification close to Gossipsub. By
contrasting this lowest layer of our refinement chain to Gossipsub, we
will be able to find and explain security issues in Gossipsub from a
refinement point of view.

\textbf{Acknowledgements}
We thank the anonymous reviewers for their thoughtful feedback and
suggestions, which helped improve the quality and clarity of this
work.

\bibliographystyle{eptcs}
\bibliography{main.bib}

\begin{thebibliography}{10}
\providecommand{\bibitemdeclare}[2]{}
\providecommand{\surnamestart}{}
\providecommand{\surnameend}{}
\providecommand{\urlprefix}{Available at }
\providecommand{\url}[1]{\texttt{#1}}
\providecommand{\href}[2]{\texttt{#2}}
\providecommand{\urlalt}[2]{\href{#1}{#2}}
\providecommand{\doi}[1]{doi:\urlalt{https://doi.org/#1}{#1}}
\providecommand{\eprint}[1]{arXiv:\urlalt{https://arxiv.org/abs/#1}{#1}}
\providecommand{\bibinfo}[2]{#2}

\bibitemdeclare{misc}{libp2p-overview}
\bibitem{libp2p-overview}
\emph{\bibinfo{title}{What is {Publish/Subscribe}}}.
\newblock
  \bibinfo{howpublished}{\url{https://docs.libp2p.io/concepts/pubsub/overview/}}.
\newblock \bibinfo{note}{Accessed 12 May 2023}.

\bibitemdeclare{incollection}{p2p-survey}
\bibitem{p2p-survey}
\bibinfo{author}{Ioannis \surnamestart Aekaterinidis\surnameend} \&
  \bibinfo{author}{Peter \surnamestart Triantafillou\surnameend}
  (\bibinfo{year}{2018}): \emph{\bibinfo{title}{Peer-to-Peer Publish-Subscribe
  Systems}}.
\newblock In: {\slshape \bibinfo{booktitle}{Encyclopedia of Database Systems,
  Second Edition}}, \doi{10.1007/978-1-4614-8265-9\_1221}.

\bibitemdeclare{inproceedings}{pastryCorrect}
\bibitem{pastryCorrect}
\bibinfo{author}{Noran \surnamestart Azmy\surnameend}, \bibinfo{author}{Stephan
  \surnamestart Merz\surnameend} \& \bibinfo{author}{Christoph \surnamestart
  Weidenbach\surnameend} (\bibinfo{year}{2016}): \emph{\bibinfo{title}{A
  Rigorous Correctness Proof for Pastry}}.
\newblock In: {\slshape \bibinfo{booktitle}{Abstract State Machines, Alloy, B,
  TLA, VDM, and {Z}}}, \doi{10.1007/978-3-319-33600-8\_5}.

\bibitemdeclare{article}{bcg88}
\bibitem{bcg88}
\bibinfo{author}{Michael~C. \surnamestart Browne\surnameend},
  \bibinfo{author}{Edmund~M. \surnamestart Clarke\surnameend} \&
  \bibinfo{author}{Orna \surnamestart Grumberg\surnameend}
  (\bibinfo{year}{1988}): \emph{\bibinfo{title}{Characterizing Finite Kripke
  Structures in Propositional Temporal Logic}}.
\newblock \doi{10.1016/0304-3975(88)90098-9}.

\bibitemdeclare{article}{castro2002scribe}
\bibitem{castro2002scribe}
\bibinfo{author}{Miguel \surnamestart Castro\surnameend},
  \bibinfo{author}{Peter \surnamestart Druschel\surnameend},
  \bibinfo{author}{A-M \surnamestart Kermarrec\surnameend} \&
  \bibinfo{author}{Antony~IT \surnamestart Rowstron\surnameend}
  (\bibinfo{year}{2002}): \emph{\bibinfo{title}{SCRIBE: A large-scale and
  decentralized application-level multicast infrastructure}}.
\newblock {\slshape \bibinfo{journal}{IEEE Journal on Selected Areas in
  communications}}, \doi{10.1109/JSAC.2002.803069}.

\bibitemdeclare{inproceedings}{acl2s11}
\bibitem{acl2s11}
\bibinfo{author}{Harsh \surnamestart Chamarthi\surnameend},
  \bibinfo{author}{Peter~C. \surnamestart Dillinger\surnameend},
  \bibinfo{author}{Panagiotis \surnamestart Manolios\surnameend} \&
  \bibinfo{author}{Daron \surnamestart Vroon\surnameend}
  (\bibinfo{year}{2011}): \emph{\bibinfo{title}{The "ACL2" Sedan Theorem
  Proving System}}.
\newblock In: {\slshape \bibinfo{booktitle}{Tools and Algorithms for the
  Construction and Analysis of Systems (TACAS)}},
  \doi{10.1007/978-3-642-19835-9_27}.

\bibitemdeclare{phdthesis}{harsh-dissertation}
\bibitem{harsh-dissertation}
\bibinfo{author}{Harsh~Raju \surnamestart Chamarthi\surnameend}
  (\bibinfo{year}{2016}): \emph{\bibinfo{title}{Interactive Non-theorem
  Disproving}}.
\newblock Ph.D. thesis, \bibinfo{school}{Northeastern University},
  \doi{10.17760/D20467205}.

\bibitemdeclare{article}{cgen}
\bibitem{cgen}
\bibinfo{author}{Harsh~Raju \surnamestart Chamarthi\surnameend},
  \bibinfo{author}{Dillinger~Peter \surnamestart C.\surnameend},
  \bibinfo{author}{Matt \surnamestart Kaufmann\surnameend} \&
  \bibinfo{author}{Panagiotis \surnamestart Manolios\surnameend}
  (\bibinfo{year}{2011}): \emph{\bibinfo{title}{Integrating testing and
  interactive theorem proving}}.
\newblock \doi{10.4204/EPTCS.70.1}.

\bibitemdeclare{article}{defdata}
\bibitem{defdata}
\bibinfo{author}{Harsh~Raju \surnamestart Chamarthi\surnameend},
  \bibinfo{author}{Dillinger~Peter \surnamestart C.\surnameend} \&
  \bibinfo{author}{Panagiotis \surnamestart Manolios\surnameend}
  (\bibinfo{year}{2014}): \emph{\bibinfo{title}{Data {Definitions} in the
  {ACL2} {Sedan}}}.
\newblock \doi{10.4204/eptcs.152.3}.

\bibitemdeclare{inproceedings}{harsh-fmcad}
\bibitem{harsh-fmcad}
\bibinfo{author}{Harsh~Raju \surnamestart Chamarthi\surnameend} \&
  \bibinfo{author}{Panagiotis \surnamestart Manolios\surnameend}
  (\bibinfo{year}{2011}): \emph{\bibinfo{title}{Automated specification
  analysis using an interactive theorem prover}}.
\newblock In \bibinfo{editor}{Per \surnamestart Bjesse\surnameend} \&
  \bibinfo{editor}{Anna \surnamestart Slobodov{\'{a}}\surnameend}, editors:
  {\slshape \bibinfo{booktitle}{International Conference on Formal Methods in
  Computer-Aided Design, {FMCAD} '11}}, \bibinfo{publisher}{{FMCAD} Inc.}, pp.
  \bibinfo{pages}{46--53}.
\newblock \urlprefix\url{http://dl.acm.org/citation.cfm?id=2157665}.

\bibitemdeclare{article}{dijkstra74}
\bibitem{dijkstra74}
\bibinfo{author}{Edsger~W. \surnamestart Dijkstra\surnameend}
  (\bibinfo{year}{1974}): \emph{\bibinfo{title}{Self-stabilizing Systems in
  Spite of Distributed Control}}.
\newblock {\slshape \bibinfo{journal}{Commun. {ACM}}},
  \doi{10.1145/361179.361202}.

\bibitemdeclare{inproceedings}{dillinger-acl2-sedan}
\bibitem{dillinger-acl2-sedan}
\bibinfo{author}{Peter~C. \surnamestart Dillinger\surnameend},
  \bibinfo{author}{Panagiotis \surnamestart Manolios\surnameend},
  \bibinfo{author}{Daron \surnamestart Vroon\surnameend} \&
  \bibinfo{author}{J.~Strother \surnamestart Moore\surnameend}
  (\bibinfo{year}{2007}): \emph{\bibinfo{title}{{ACL2s}: ``The {ACL2}
  Sedan''}}.
\newblock In: {\slshape \bibinfo{booktitle}{Proceedings of the 7th Workshop on
  User Interfaces for Theorem Provers (UITP 2006)}},
  \doi{10.1016/j.entcs.2006.09.018}.

\bibitemdeclare{inproceedings}{hawblitzel2015ironfleet}
\bibitem{hawblitzel2015ironfleet}
\bibinfo{author}{Chris \surnamestart Hawblitzel\surnameend},
  \bibinfo{author}{Jon \surnamestart Howell\surnameend}, \bibinfo{author}{Manos
  \surnamestart Kapritsos\surnameend}, \bibinfo{author}{Jacob~R \surnamestart
  Lorch\surnameend}, \bibinfo{author}{Bryan \surnamestart Parno\surnameend},
  \bibinfo{author}{Michael~L \surnamestart Roberts\surnameend},
  \bibinfo{author}{Srinath \surnamestart Setty\surnameend} \&
  \bibinfo{author}{Brian \surnamestart Zill\surnameend} (\bibinfo{year}{2015}):
  \emph{\bibinfo{title}{IronFleet: proving practical distributed systems
  correct}}.
\newblock In: {\slshape \bibinfo{booktitle}{Proceedings of the 25th Symposium
  on Operating Systems Principles}}, \doi{10.1145/2815400.2815428}.

\bibitemdeclare{article}{alloy}
\bibitem{alloy}
\bibinfo{author}{Daniel \surnamestart Jackson\surnameend}
  (\bibinfo{year}{2019}): \emph{\bibinfo{title}{Alloy: a language and tool for
  exploring software designs}}.
\newblock \doi{10.1145/3338843}.

\bibitemdeclare{book}{acl2-car}
\bibitem{acl2-car}
\bibinfo{author}{Matt \surnamestart Kaufmann\surnameend},
  \bibinfo{author}{Panagiotis \surnamestart Manolios\surnameend} \&
  \bibinfo{author}{J~Strother \surnamestart Moore\surnameend}
  (\bibinfo{year}{2000}): \emph{\bibinfo{title}{Computer-Aided Reasoning: An
  Approach}}.
\newblock \bibinfo{publisher}{Kluwer Academic Publishers},
  \doi{10.1007/978-1-4615-4449-4}.

\bibitemdeclare{book}{acl2-acs}
\bibitem{acl2-acs}
\bibinfo{author}{Matt \surnamestart Kaufmann\surnameend},
  \bibinfo{author}{Panagiotis \surnamestart Manolios\surnameend} \&
  \bibinfo{author}{J~Strother \surnamestart Moore\surnameend}
  (\bibinfo{year}{2000}): \emph{\bibinfo{title}{Computer-Aided Reasoning: Case
  Studies}}.
\newblock \bibinfo{publisher}{Kluwer Academic Publishers},
  \doi{10.1007/978-1-4757-3188-0}.

\bibitemdeclare{misc}{acl2-web}
\bibitem{acl2-web}
\bibinfo{author}{Matt \surnamestart Kaufmann\surnameend} \&
  \bibinfo{author}{J~Strother \surnamestart Moore\surnameend}
  (\bibinfo{year}{2022}): \emph{\bibinfo{title}{{ACL2 homepage}}}.
\newblock \urlprefix\url{https://www.cs.utexas.edu/users/moore/acl2/}.

\bibitemdeclare{inproceedings}{KumarGWS}
\bibitem{KumarGWS}
\bibinfo{author}{Ankit \surnamestart Kumar\surnameend}, \bibinfo{author}{Max
  \surnamestart von Hippel\surnameend}, \bibinfo{author}{Panagiotis
  \surnamestart Manolios\surnameend} \& \bibinfo{author}{Cristina \surnamestart
  Nita{-}Rotaru\surnameend} (\bibinfo{year}{2023}):
  \emph{\bibinfo{title}{Verification of GossipSub in ACL2s}}.
\newblock In: {\slshape \bibinfo{booktitle}{International Workshop on the
  {ACL2} Theorem Prover and Its Applications}}, \doi{10.4204/EPTCS.393.10}.

\bibitemdeclare{inproceedings}{KumarGossip}
\bibitem{KumarGossip}
\bibinfo{author}{Ankit \surnamestart Kumar\surnameend}, \bibinfo{author}{Max
  \surnamestart von Hippel\surnameend}, \bibinfo{author}{Panagiotis
  \surnamestart Manolios\surnameend} \& \bibinfo{author}{Cristina \surnamestart
  Nita{-}Rotaru\surnameend} (\bibinfo{year}{2024}):
  \emph{\bibinfo{title}{Formal Model-Driven Analysis of Resilience of GossipSub
  to Attacks from Misbehaving Peers}}.
\newblock In: {\slshape \bibinfo{booktitle}{{IEEE} Symposium on Security and
  Privacy, {SP} 2024, San Francisco, CA, USA, May 19-23, 2024}},
  \doi{10.1109/SP54263.2024.00017}.

\bibitemdeclare{inproceedings}{ankit-mpmt}
\bibitem{ankit-mpmt}
\bibinfo{author}{Ankit \surnamestart Kumar\surnameend} \&
  \bibinfo{author}{Panagiotis \surnamestart Manolios\surnameend}
  (\bibinfo{year}{2021}): \emph{\bibinfo{title}{Mathematical Programming Modulo
  Strings}}.
\newblock In: {\slshape \bibinfo{booktitle}{Formal Methods in Computer Aided
  Design, {FMCAD}}}, \doi{10.34727/2021/ISBN.978-3-85448-046-4\_36}.

\bibitemdeclare{misc}{floodsubref}
\bibitem{floodsubref}
\bibinfo{author}{Ankit \surnamestart Kumar\surnameend} \&
  \bibinfo{author}{Panagiotis \surnamestart Manolios\surnameend}
  (\bibinfo{year}{2025}): \emph{\bibinfo{title}{Proof of Refinement of
  Floodsub}}.
\newblock \urlprefix\url{https://github.com/ankitku/FloodsubRef}.
\newblock \bibinfo{note}{In submission}.

\bibitemdeclare{article}{lamport2002specifying}
\bibitem{lamport2002specifying}
\bibinfo{author}{Leslie \surnamestart Lamport\surnameend}
  (\bibinfo{year}{2002}): \emph{\bibinfo{title}{Specifying systems: the TLA+
  language and tools for hardware and software engineers}}.

\bibitemdeclare{phdthesis}{pete-dissertation}
\bibitem{pete-dissertation}
\bibinfo{author}{Panagiotis \surnamestart Manolios\surnameend}
  (\bibinfo{year}{2001}): \emph{\bibinfo{title}{Mechanical Verification of
  Reactive Systems}}.
\newblock Ph.D. thesis, \bibinfo{school}{The University of Texas at Austin,
  Department of Computer Sciences, Austin TX}.

\bibitemdeclare{inproceedings}{DBLP:conf/date/ManoliosS04}
\bibitem{DBLP:conf/date/ManoliosS04}
\bibinfo{author}{Panagiotis \surnamestart Manolios\surnameend} \&
  \bibinfo{author}{Sudarshan~K. \surnamestart Srinivasan\surnameend}
  (\bibinfo{year}{2004}): \emph{\bibinfo{title}{Automatic Verification of
  Safety and Liveness for XScale-Like Processor Models Using {WEB}
  Refinements}}.
\newblock In: {\slshape \bibinfo{booktitle}{Design, Automation and Test in
  Europe Conference and Exposition, {DATE}}}, \doi{10.1109/DATE.2004.1268844}.

\bibitemdeclare{article}{DBLP:journals/tvlsi/ManoliosS08}
\bibitem{DBLP:journals/tvlsi/ManoliosS08}
\bibinfo{author}{Panagiotis \surnamestart Manolios\surnameend} \&
  \bibinfo{author}{Sudarshan~K. \surnamestart Srinivasan\surnameend}
  (\bibinfo{year}{2008}): \emph{\bibinfo{title}{A Refinement-Based
  Compositional Reasoning Framework for Pipelined Machine Verification}}.
\newblock {\slshape \bibinfo{journal}{{IEEE} Trans. Very Large Scale Integr.
  Syst.}}, \doi{10.1109/TVLSI.2008.918120}.

\bibitemdeclare{inproceedings}{ManoliosVroon03}
\bibitem{ManoliosVroon03}
\bibinfo{author}{Panagiotis \surnamestart Manolios\surnameend} \&
  \bibinfo{author}{Daron \surnamestart Vroon\surnameend}
  (\bibinfo{year}{2003}): \emph{\bibinfo{title}{Algorithms for Ordinal
  Arithmetic}}.
\newblock In: {\slshape \bibinfo{booktitle}{Conference on Automated Deduction
  {CADE}}}, \doi{10.1007/978-3-540-45085-6_19}.

\bibitemdeclare{inproceedings}{ManoliosVroon04}
\bibitem{ManoliosVroon04}
\bibinfo{author}{Panagiotis \surnamestart Manolios\surnameend} \&
  \bibinfo{author}{Daron \surnamestart Vroon\surnameend}
  (\bibinfo{year}{2004}): \emph{\bibinfo{title}{Integrating Reasoning about
  Ordinal Arithmetic into {ACL2}}}.
\newblock In: {\slshape \bibinfo{booktitle}{Formal Methods in Computer-Aided
  Design {FMCAD}}}, \bibinfo{series}{LNCS},
  \bibinfo{publisher}{Springer--Verlag}, \doi{10.1007/978-3-540-30494-4_7}.

\bibitemdeclare{article}{MV05}
\bibitem{MV05}
\bibinfo{author}{Panagiotis \surnamestart Manolios\surnameend} \&
  \bibinfo{author}{Daron \surnamestart Vroon\surnameend}
  (\bibinfo{year}{2005}): \emph{\bibinfo{title}{{Ordinal Arithmetic: Algorithms
  and Mechanization}}}.
\newblock {\slshape \bibinfo{journal}{Journal of Automated Reasoning}},
  \doi{10.1007/s10817-005-9023-9}.

\bibitemdeclare{inproceedings}{ccg}
\bibitem{ccg}
\bibinfo{author}{Panagiotis \surnamestart Manolios\surnameend} \&
  \bibinfo{author}{Daron \surnamestart Vroon\surnameend}
  (\bibinfo{year}{2006}): \emph{\bibinfo{title}{Termination Analysis with
  Calling Context Graphs}}.
\newblock In: {\slshape \bibinfo{booktitle}{Computer Aided Verification
  {CAV}}}, \doi{10.1007/11817963\_36}.

\bibitemdeclare{misc}{manolios2023reasoning}
\bibitem{manolios2023reasoning}
\bibinfo{author}{Pete \surnamestart Manolios\surnameend}
  (\bibinfo{year}{2023}): \emph{\bibinfo{title}{Reasoning About Programs}}.
\newblock
  \urlprefix\url{https://www.ccs.neu.edu/home/pete/courses/Logic-and-Computation/2023-Fall/lectures.html}.
\newblock \bibinfo{note}{Lecture notes, Northeastern University, CS 2800: Logic
  and Computation, Accessed 21 Apr 2025}.

\bibitemdeclare{inproceedings}{milner71}
\bibitem{milner71}
\bibinfo{author}{Robin \surnamestart Milner\surnameend} (\bibinfo{year}{1971}):
  \emph{\bibinfo{title}{An Algebraic Definition of Simulation Between
  Programs}}.
\newblock In: {\slshape \bibinfo{booktitle}{Proceedings of the 2nd
  International Joint Conference on Artificial Intelligence}}.
\newblock \urlprefix\url{http://ijcai.org/Proceedings/71/Papers/044.pdf}.

\bibitemdeclare{inproceedings}{padon-ivy}
\bibitem{padon-ivy}
\bibinfo{author}{Oded \surnamestart Padon\surnameend},
  \bibinfo{author}{Kenneth~L. \surnamestart McMillan\surnameend},
  \bibinfo{author}{Aurojit \surnamestart Panda\surnameend},
  \bibinfo{author}{Mooly \surnamestart Sagiv\surnameend} \&
  \bibinfo{author}{Sharon \surnamestart Shoham\surnameend}
  (\bibinfo{year}{2016}): \emph{\bibinfo{title}{Ivy: safety verification by
  interactive generalization}}.
\newblock \doi{10.1145/2908080.2908118}.

\bibitemdeclare{inproceedings}{pastry-p2p-location}
\bibitem{pastry-p2p-location}
\bibinfo{author}{Antony I.~T. \surnamestart Rowstron\surnameend} \&
  \bibinfo{author}{Peter \surnamestart Druschel\surnameend}
  (\bibinfo{year}{2001}): \emph{\bibinfo{title}{Pastry: Scalable, Decentralized
  Object Location, and Routing for Large-Scale Peer-to-Peer Systems}}.
\newblock In: {\slshape \bibinfo{booktitle}{International Conference on
  Distributed Systems Platforms}}, \doi{10.1007/3-540-45518-3\_18}.

\bibitemdeclare{inproceedings}{derecho}
\bibitem{derecho}
\bibinfo{author}{Kumar \surnamestart Shivam\surnameend},
  \bibinfo{author}{Vishnu \surnamestart Paladugu\surnameend} \&
  \bibinfo{author}{Yanhong~A. \surnamestart Liu\surnameend}
  (\bibinfo{year}{2023}): \emph{\bibinfo{title}{Specification and Runtime
  Checking of Derecho, {A} Protocol for Fast Replication for Cloud Services}}.
\newblock \doi{10.1145/3584684.3597275}.

\bibitemdeclare{misc}{gsreadme}
\bibitem{gsreadme}
\bibinfo{author}{Dimitris \surnamestart Vyzovitis\surnameend}:
  \emph{\bibinfo{title}{gossipsub: An extensible baseline pubsub protocol}}.
\newblock
  \bibinfo{howpublished}{\url{https://github.com/libp2p/specs/blob/master/pubsub/gossipsub/README.md}}.
\newblock \bibinfo{note}{Accessed 28 Nov 2022}.

\bibitemdeclare{misc}{floodsub}
\bibitem{floodsub}
\bibinfo{author}{Dimitris \surnamestart Vyzovitis\surnameend}
  (\bibinfo{year}{2020}): \emph{\bibinfo{title}{GossipSub~v1.0: An extensible
  baseline pubsub protocol}}.
\newblock
  \bibinfo{howpublished}{\url{https://github.com/libp2p/specs/blob/master/pubsub/gossipsub/gossipsub-v1.0-old.md}}.
\newblock \bibinfo{note}{Accessed 23 Jan 2025}.

\bibitemdeclare{inproceedings}{invariant-discovery-game}
\bibitem{invariant-discovery-game}
\bibinfo{author}{Andrew~T. \surnamestart Walter\surnameend},
  \bibinfo{author}{Benjamin \surnamestart Boskin\surnameend},
  \bibinfo{author}{Seth \surnamestart Cooper\surnameend} \&
  \bibinfo{author}{Panagiotis \surnamestart Manolios\surnameend}
  (\bibinfo{year}{2019}): \emph{\bibinfo{title}{Gamification of Loop-Invariant
  Discovery from Code}}.
\newblock In: {\slshape \bibinfo{booktitle}{Proceedings of the Seventh {AAAI}
  Conference on Human Computation and Crowdsourcing, {HCOMP}}},
  \doi{10.1609/HCOMP.V7I1.5277}.

\bibitemdeclare{inproceedings}{enumerative-data-types}
\bibitem{enumerative-data-types}
\bibinfo{author}{Andrew~T. \surnamestart Walter\surnameend},
  \bibinfo{author}{David~A. \surnamestart Greve\surnameend} \&
  \bibinfo{author}{Panagiotis \surnamestart Manolios\surnameend}
  (\bibinfo{year}{2022}): \emph{\bibinfo{title}{Enumerative Data Types with
  Constraints}}.
\newblock In: {\slshape \bibinfo{booktitle}{Formal Methods in Computer-Aided
  Design, {FMCAD}}}, \doi{10.34727/2022/ISBN.978-3-85448-053-2\_25}.

\bibitemdeclare{inproceedings}{acl2-workshop-checker-paper}
\bibitem{acl2-workshop-checker-paper}
\bibinfo{author}{Andrew~T. \surnamestart Walter\surnameend},
  \bibinfo{author}{Ankit \surnamestart Kumar\surnameend} \&
  \bibinfo{author}{Panagiotis \surnamestart Manolios\surnameend}
  (\bibinfo{year}{2023}): \emph{\bibinfo{title}{Proving Calculational Proofs
  Correct}}.
\newblock In: {\slshape \bibinfo{booktitle}{Proceedings of the 18th
  International Workshop on the {ACL2} Theorem Prover and Its Applications}},
  \doi{10.4204/EPTCS.393.11}.

\bibitemdeclare{inproceedings}{acl2s-systems-programming}
\bibitem{acl2s-systems-programming}
\bibinfo{author}{Andrew~T. \surnamestart Walter\surnameend} \&
  \bibinfo{author}{Panagiotis \surnamestart Manolios\surnameend}
  (\bibinfo{year}{2022}): \emph{\bibinfo{title}{{ACL2s} Systems Programming}}.
\newblock In: {\slshape \bibinfo{booktitle}{Workshop on the {ACL2} Theorem
  Prover and its Applications}}, \doi{10.4204/EPTCS.359.12}.

\bibitemdeclare{article}{chordZave}
\bibitem{chordZave}
\bibinfo{author}{Pamela \surnamestart Zave\surnameend} (\bibinfo{year}{2012}):
  \emph{\bibinfo{title}{Using lightweight modeling to understand chord}}.
\newblock {\slshape \bibinfo{journal}{Comput. Commun. Rev.}},
  \doi{10.1145/2185376.2185383}.

\end{thebibliography}
\end{document}